\documentclass[12pt, letter]{article}

\usepackage{amssymb,amsfonts,amsmath}
\newcommand{\be}{\begin{equation}}
\newcommand{\ee}{\end{equation}}
\newcommand{\ba}{\begin{array}}
\newcommand{\ea}{\end{array}}
\newcommand{\p}{\partial}

\newcommand{\ds}{\displaystyle}
\newtheorem{theo}{Theorem}

\newtheorem{cor}{Corollary}
\newtheorem{prop}{Proposition}

\begin{document}%\Title{preprint arXiv:0806.0177 [math-ph]}
\thispagestyle{empty}
\title
%[Why nonlocal recursion operators produce local symmetries]
{\protect\vspace*{-20mm}{\bf Infinite hierarchies of nonlocal symmetries\\
of the Chen--Kontsevich--Schwarz type\\
for the oriented associativity equations}
%\thanks{accepted for publication in {\em J. Phys. A: Math. Theor.}}
\protect\vspace{-3mm}}
\author{{\sc A. Sergyeyev}\\[1mm]
%\address{
%\normalsize
Mathematical Institute, Silesian University in Opava,\\
Na Rybn\'\i{}\v{c}ku 1, 746\,01 Opava, Czech Republic\\ E-mail: {\tt
Artur.Sergyeyev@math.slu.cz}}
\date{June 14, 2009}
%\date{}
\maketitle
%\maketitle %\protect\vspace{-10mm}
\begin{abstract}\protect\vspace{-12mm}

We construct infinite hierarchies of nonlocal higher symmetries
for the oriented associativity equations
using solutions of associated vector and scalar spectral problems.
The symmetries in question generalize those found
by Chen, Kontsevich and Schwarz \cite{cks}
for the WDVV equations. As a byproduct, we obtain a Darboux-type transformation
and a (conditional) B\"acklund transformation for the oriented associativity equations.\looseness=-1
\end{abstract}
\vspace{-9mm}
\section*{Introduction}
%\label{sec:indroduction}
%\setcounter{equation}{0}
The Witten--Dijkgraaf--Verlinde--Verlinde (WDVV) equations \cite{wi,dvv},
and the related geometric structures, in particular,
the Frobenius manifolds %\cite{dub,dub2,konman,dub1,manin_bk, hertling_bk},
\cite{dub}--\cite{hertling_bk},
%and the $F$-manifolds \cite{hm,man,lm},
have attracted
considerable attention because
of their manifold applications in physics and mathematics.
\looseness=-1

More recently, the oriented
associativity equations, a generalization of the WDVV equations, %see (\ref{ae}) below,
and the related geometric structures, $F$-manifolds,
see e.g.\ %\cite{manin_bk, hertling_bk, man, hm, lm, lp, man05, kon, k3},
\cite{manin_bk}--\cite{k2},
have also become
a subject of intense research.
These equations have first appeared in \cite[Proposition 2.3]{dub}
as the equations for the displacement vector.
The oriented
associativity equations describe {\em inter alia} isoassociative quantum
deformations of commutative associative algebras \cite{kon, k3},
cf.\ also \cite{km}--\cite{k2}.
%\cite{km, km2, k1, k2}.
\looseness=-1

The oriented associativity equations (\ref{ae}) admit the gradient reduction (\ref{gr}) to the
``usual" associativity equations (\ref{wdvv}).
%, i.e.,
%which is nothing but the WDVV equations
%stripped of the quasihomogeneity condition and
%the condition (\ref{etac}) expressing existence of
%the unit element in the related associative algebra.
Equations (\ref{wdvv}) and the so-called Hessian reduction (see \cite{m5, kito, m6, m_geo3, dub5} and \cite{kon})
of the oriented associativity equations naturally arise in
%topological sigma-models and
topological 2D gravity \cite{wi,wi2}, singularity theory and complex geometry
(see e.g.\ \cite{hertling_bk, man}),
and in differential geometry and theory of integrable systems,
%theory of hydrodynamic-type systems,
% and were extensively studied in this context,
see \cite{hertling_bk}, \cite{mbook}--\cite{boy}
%\cite{hertling_bk, mbook, m_geo2, m_geo3, m_geo4, m_geo5, fer, str,
%david, david2, mvp, mvp2, boy}
and references therein.\looseness=-1

There is a considerable body of work on the
symmetry properties of the WDVV equations,
see e.g.\  \cite{mm, gmp, giv, cks, conte} for the point symmetries of the WDVV and
generalized (in the sense of \cite{mor} and references therein) WDVV equations,
\cite{dub2,wit,hoev,riley} and references therein
for finite symmetries, B\"acklund transformations and dualities, and
\cite{mbook, mf1,mf2,mf3} and references therein for the higher symmetries and (bi-)Hamiltonian
structures for the WDVV equations, and also for (\ref{wdvv}), in three and four independent variables.
Although the approach of \cite{mbook, mf1,mf2,mf3} in principle could \cite{mbook} be
generalized to the WDVV equations in more than four independent variables,
this was not done yet. Nevertheless, in \cite{giv, cks} infinite sets of {\em nonlocal}
higher symmetries for the WDVV equations were found.
To the best of our knowledge, higher (or generalized \cite{olv_eng2})
symmetries
of the {\em oriented} associativity equations
in arbitrary dimension were never fully explored.\looseness=-1

The goal of the present paper is to construct nonlocal
higher symmetries for the oriented associativity equations equations
using, in analogy with \cite{cks}, the solutions of auxiliary spectral problems.
We show that the very solutions of the vector spectral problem (\ref{zcr}), either {\em per se}
or multiplied by a suitably chosen solution of the scalar spectral problem (\ref{ssp_oae})
with the opposite sign of the spectral parameter,
indeed {\em are} (infinitesimal) %nonlocal
symmetries for the oriented associativity equations (\ref{ae}), see
Theorem~\ref{sp} below for details. The fact that solutions of
(\ref{zcr}) are symmetries for (\ref{ae}) is quite unusual in
itself, as symmetries typically turn out to be quadratic
\cite{konbook} rather than linear in solutions of auxiliary linear
problems.\looseness=-1

Upon performing the gradient reduction to the ``usual" associativity equations (\ref{wdvv})
we reproduce the results of \cite{cks}, see Corollaries~\ref{sp1} and \ref{oae_new_sym2wdvv} below.
However, not all nonlocal symmetries from Theorem~\ref{sp} survive the gradient reduction
and yield symmetries for (\ref{wdvv}), see Corollary~\ref{sp1} and the subsequent discussion.
\looseness=-1

Expanding solutions of the spectral problems into formal
Taylor series in the spectral parameter
yields infinite hierarchies of nonlocal higher symmetries
for (\ref{ae}) and (\ref{wdvv}),
see Corollaries~\ref{oae_new_sym2}~and~\ref{oae_new_sym2wdvv} below.\looseness=-1
\looseness=-1

Finally, as a byproduct, we obtain a Darboux-type transformation
and some B\"acklund-type transformations relating the
solutions of ``usual" and oriented associativity equations, see Proposition~\ref{dt2} and
Corollaries~\ref{bt} and \ref{sol_oae} for further
details. These transformations, as well as the nonlocal symmetries discussed above, could possibly yield
new solutions for the oriented and ``usual" associativity equations.\looseness=-1

\section{Preliminaries}
Let the Greek indices $\alpha,\beta,\gamma,\dots$
(except for $\lambda,\mu,\eta,\zeta,\sigma,\tau,\chi,\phi,\psi$)
run from 1 to $n$, where $n$ is a fixed natural number, and summation over the repeated indices
be understood unless otherwise explicitly stated. In what follows we also assume that
all functions under study are sufficiently smooth for all necessary derivatives to exist.
\looseness=-1

%The system of equations for $K^\alpha$ obtained by substitution of (\ref{cs})
%into (\ref{ae0}), that is,

%\newpage
Consider the {\em oriented associativity equations}, see e.g.\ \cite{hm}--\cite{kon},
for the structure ``constants" $c_{\alpha\beta}^\delta(x^1,\dots,\allowbreak x^n)$ of a commutative
($c^\alpha_{\nu\rho}=c^\alpha_{\rho\nu}$) algebra:
\begin{eqnarray}
c^\nu_{\alpha\rho}c^{\rho}_{\beta\gamma}=c^\nu_{\rho\gamma}c^\rho_{\alpha\beta},\label{ae0}\\
\frac{\p c^{\alpha}_{\beta\gamma}}{\p x^\rho}=\frac{\p
c^{\alpha}_{\rho\gamma}}{\p x^\beta}.\label{dif}
\end{eqnarray}
The condition (\ref{ae0}) means that the algebra in question is associative,
and (\ref{dif}) means that we consider isoassociative \cite{kon} quantum deformations
of the algebra in question.

The oriented associativity equations (\ref{ae0}), (\ref{dif}) can be written as compatibility conditions
of the Gauss--Manin equations, see e.g.\ \cite{dub2,manin_bk,kon}, for a scalar function $\chi(\lambda)$
(for the sake of brevity we shall often omit below the dependence on $x^1,\dots,x^n$):
\begin{equation}
\label{zcrchi_oae}
  \frac{\partial^2\chi(\lambda)}{\partial x^{\alpha} \partial x^{\gamma}}= \lambda
%\frac{\partial^2 K^\nu}{\partial x^{\alpha} \partial x^{\gamma}}
c^\nu_{\alpha\gamma}
 \frac{\partial\chi(\lambda)}{\partial x^{\nu}}.
\end{equation}
Here $\lambda$ is the spectral parameter.
These equations have a very interesting interpretation,
with $\chi$ playing the role of a wave function,
in the context of quantum deformations of associative algebras \cite{kon}.

%For the future references, we shall also need the formal adjoint of (\ref{zcrchi_oae}), viz.,
%\begin{equation}
%\label{zcrchibar}
%  \frac{\partial^2\bar\chi}{\partial x^{\alpha} \partial x^{\gamma}}= -\lambda
%\frac{\partial^2 K^\nu}{\partial x^{\alpha} \partial x^{\gamma}}
% \frac{\partial\bar\chi}{\partial x^{\nu}}\/.
%\end{equation}

We also have a
zero-curvature representation for (\ref{ae0}), (\ref{dif}) of the form
(see e.g.\ \cite{dub1, mor})
\be\label{zcr0}
\frac{\p \psi^\alpha(\lambda)}{\p x^\beta}=\lambda c^\alpha_{\beta\gamma}\psi^\gamma(\lambda),
%\frac{\p^2 K^\alpha}{\p x^\beta\p x^\gamma}\psi^\gamma,
\ee
where we again omit, for the sake of brevity, the dependence of $\psi^\alpha$ on $x^1,\dots,x^n$.
In other words, Eqs.(\ref{ae0}), (\ref{dif}) are precisely the
compatibility conditions for (\ref{zcr0}).
The quantities $\psi^\alpha(\lambda)$ are nothing but the components of a generic vector field
which is covariantly constant (in other terminology, parallel or flat) with respect to the covariant derivative
associated with the one-parametric
family of flat connections $-\lambda c^\alpha_{\nu\rho}$. \looseness=-1

Upon introducing the quantities $\phi_\alpha(\lambda)=\p \chi(\lambda)/\p x^\alpha$
%, we readily see that
Eq.(\ref{zcrchi_oae}) can be written in the first-order form as
\be\label{zcr_phi_oae}
\p\phi_\alpha(\lambda)/\p x^\beta=\lambda c_{\alpha\beta}^\delta \phi_\delta(\lambda).
\ee
Quite obviously, the spectral problem (\ref{zcr_phi_oae}) is, up to the change of sign of $\lambda$,
adjoint to (\ref{zcr0}).

Now let $\chi^\alpha(\lambda)$, $\alpha=1,\dots,n$,
be the solutions of (\ref{zcrchi_oae}) normalized by the condition
\[
\chi^\alpha(\lambda)|_{\lambda=0}=x^\alpha.
\]
It is well known (see e.g.\ \cite{dub2, hertling_bk, kon}) that $\chi^\alpha$
are nothing but flat coordinates for the one-parameter
family $\lambda c^\alpha_{\nu\kappa}$ of flat connections (the flatness readily follows
from (\ref{ae0}) and (\ref{dif})).

Following \cite{dub,dub2}, we can represent $\chi^\alpha$ in the form
\be\label{chiaexp}
\chi^\alpha(\lambda)=x^\alpha+\lambda K^\alpha+O(\lambda^2),
\ee
where $K^\alpha=K^\alpha(x^1,\dots,x^n)$ is the so-called displacement vector.
Plugging (\ref{chiaexp}) into (\ref{zcrchi_oae})
and restricting our attention to the terms linear in $\lambda$
yields
\be\label{cs}
c^\alpha_{\beta\gamma}=\ds\frac{\p^2 K^\alpha}{\p x^\beta\p x^\gamma}.
\ee

The ansatz (\ref{cs}) automatically solves (\ref{dif}), and (\ref{ae0}) boils down to the
overdetermined system
\be\label{ae}
\ds\frac{\p^2 K^\nu}{\p x^\alpha\p x^\rho}\frac{\p^2 K^\rho}{\p x^\beta\p x^\gamma}
=\frac{\p^2 K^\rho}{\p x^\alpha\p x^\beta} \frac{\p^2 K^\nu}{\p x^\rho\p x^\gamma}
\ee
for $K^\alpha$. In what follows we shall refer to this system
as to the oriented associativity equations just as we referred to
(\ref{ae0}), (\ref{dif}), as, in combination with (\ref{cs}),
Eq.(\ref{ae}) is equivalent to (\ref{ae0}), (\ref{dif}) provided $c^\alpha_{\nu\rho}=c^\alpha_{\rho\nu}$.

%we see that Eqs.(\ref{ae}) are nothing but the associativity conditions
%for a commutative algebra with the structure ``constants"
%(\ref{cs}), see e.g.\ \cite{dub,dub1,lm, kon}.\looseness=-2
% of a commutative
%Following \cite{dub,dub1,giv0,kon,lm} consider
%the associativity equations
%\be\label{ae0}
%c^\nu_{ar}c^\rho_{\beta\gamma}=c^\rho_{\alpha\beta}c^\nu_{rc},
%\ee

%Eqs.(\ref{ae}) along with the obvious symmetry
%$c^\alpha_{\beta\gamma}=c^\alpha_{cb}$ that follows from (\ref{cs})
%imply $c^\alpha_{\beta\gamma}$ are structure constants of an associative commutative algebra $\mathcal{A}$.
%The commutativity of the algebra is ensured by (\ref{cs}), as we obviously have
%\[
%c^\alpha_{\beta\gamma}=c^\alpha_{cb}
%\]
%by virtue of (\ref{cs}).

Of course, the equations obtained by plugging (\ref{cs}) into (\ref{zcr0}), that is,
\be\label{zcr}
\frac{\p \psi^\alpha(\lambda)}{\p x^\beta}=\lambda \frac{\p^2 K^\alpha}{\p x^\beta\p x^\gamma}\psi^\gamma(\lambda),
\ee
provide a zero-curvature representation for (\ref{ae}).

Likewise, plugging (\ref{cs}) into (\ref{zcrchi_oae}) yields a scalar spectral problem for (\ref{ae}):
\be\label{ssp_oae}
  \frac{\partial^2\chi(\lambda)}{\partial x^{\alpha} \partial x^{\gamma}}= \lambda
%\frac{\partial^2 K^\nu}{\partial x^{\alpha} \partial x^{\gamma}}
\frac{\p^2 K^\nu}{\p x^\alpha\p x^\gamma}
 \frac{\partial\chi(\lambda)}{\partial x^{\nu}}.
\ee

%However, such a relationship does exist for the gradient reduction (\ref{wdvv})
%of (\ref{ae}), see Eq.(\ref{grzcr}) below.

\section{Darboux-type transformation for oriented\\ associativity
equations}
It is well known, see e.g.\ \cite[Appendix B]{dub2}, \cite{mm} and references therein, that
%for the (generalized)
%WDVV equations
there exist changes of variables that leave the second derivatives
of the prepotential unchanged and map solutions of the (generalized)
WDVV equations into new solutions. Quite interestingly, there is a change of variables of this kind that
involves \cite{mm} a solution of the spectral problem (\ref{zcr0}), and thus can be thought of
as a Darboux-type transformation.\looseness=-1

It turns out that this transformation is readily generalized to the oriented
associativity equations. Namely, the following assertion holds.
\begin{prop}\label{dt2}
Let $K^\alpha$
satisfy (\ref{ae}) and $\psi^\alpha(\lambda)$ solve the spectral problem (\ref{zcr}).
\looseness=-1
%\be\label{sp}
%\ds\frac{\p\psi^i}{\p a^j}=\lambda c^i_{jk}\psi^k. \ee
Suppose that $$\det\p\psi^\alpha(\lambda)/\p x^\beta\not\equiv~0,$$
introduce new independent variables
\be\label{asnew2} \tilde
x^\alpha=\psi^\alpha(\lambda),
\ee
and define (locally) new dependent variables
$\tilde K^\beta$ by the formulas
\be\label{csnew2a} \frac{\p\tilde
K^\beta}{\p\tilde{x}^\gamma}=\frac{\p K^\beta}{\p x^\gamma}.
\ee

Then
\be\label{csnew3} \tilde
c^\alpha_{\beta\gamma}=\frac{\p^2 \tilde K^\alpha}{\p\tilde{x}^\beta \p\tilde{x}^\gamma},
\ee
where
$\tilde K^\alpha=\tilde K^\alpha(\tilde x)$ are determined from
(\ref{csnew2a}),
satisfy
\be\label{ae2a} \tilde{c}^\alpha_{\beta\gamma}
\tilde{c}^\gamma_{\rho\nu}=\tilde{c}^\alpha_{\nu\gamma}
\tilde{c}^\gamma_{\rho\beta}.
\ee
\end{prop}

{\em Proof.} First of all, we need to show that (\ref{csnew2a}) is
well-defined, i.e., that there exist, at least locally, the functions $\tilde
K^\alpha(\tilde x)$ such that (\ref{csnew2a}) holds.
Quite clearly, this amounts to proving that we have
\be\label{sym0a}
\frac{\p^2\tilde K^\alpha}{\p\tilde{x}^\beta \p\tilde{x}^\gamma}=\frac{\p^2\tilde
K^\alpha}{\p\tilde{x}^\gamma \p\tilde{x}^\beta}, \ee
or equivalently (by virtue of
(\ref{csnew2a})),
\be\label{csnewsyma} \ds\frac{\p^2 K^\alpha}{\p
x^\beta\p\tilde{x}^\gamma}=\ds\frac{\p^2 K^\alpha}{\p x^\gamma\p\tilde{x}^\beta}.
\ee
From
(\ref{csnew2a}) we have
\be\label{css2} \ds\frac{\p^2 K^\alpha}{\p
x^\beta\p\tilde{x}^\gamma} =\ds\frac{\p^2 K^\alpha}{\p x^\beta\p x^\varepsilon}\frac{\p x^\varepsilon}{\p
\tilde{x}^\gamma}=c^\alpha_{\beta\varepsilon}\frac{\p x^\varepsilon}{\p \psi^\gamma}.
\ee
Hence,
\be\label{ccc} \ds\frac{\p^2 K^\alpha}{\p x^\beta\p\tilde{x}^\gamma}
-\ds\frac{\p^2 K^\alpha}{\p x^\gamma\p\tilde{x}^\beta}=\left(c^{\alpha}_{\beta\varepsilon}\frac{\p
x^\varepsilon}{\p \psi^\gamma}-c^{\alpha}_{\gamma\varepsilon}\frac{\p x^\varepsilon}{\p \psi^\beta}\right).
\ee
We want to show that the expression on the left-hand side of (\ref{ccc}) vanishes. But this
is equivalent to vanishing of the following quantity:
\[
B^\alpha_{\rho\nu}= \ds \left(c^{\alpha}_{\beta\varepsilon} \frac{\p x^\varepsilon}{\p
\psi^\gamma}-c^{\alpha}_{\gamma\varepsilon}\frac{\p x^\varepsilon}{\p \psi^\beta}\right) \frac{\p
\psi^\gamma}{\p x^\rho}\frac{\p \psi^\beta}{\p x^\nu}.
\]
Using the obvious identity
\[
\frac{\p \psi^\gamma}{\p x^\rho}\frac{\p x^\rho}{\p \psi^\nu}=\delta^\gamma_\nu,
\]
where $\delta^\gamma_\nu$ is the Kronecker delta, we find that
\[
B^\alpha_{\rho\nu}=\left(c^{\alpha}_{\beta\rho}\frac{\p \psi^\beta}{\p x^\nu}-c^{\alpha}_{\gamma\nu}\frac{\p
\psi^\gamma}{\p x^\rho}\right).
\]
Now using %(\ref{sp}) and
(\ref{zcr0}) we obtain
\[
B^\alpha_{\rho\nu}=\lambda\left(c^{\alpha}_{\gamma\rho}c^\gamma_{\nu\kappa}-c^{\alpha}_{\gamma\nu}c^\gamma_{\rho\kappa}\right)\psi^\kappa
\stackrel{(\ref{sp})}{=}0,
\]
and thus
\[
\ds\frac{\p^2 K^\alpha}{\p x^\beta\p\tilde{x}^\gamma} -\ds\frac{\p^2 K^\alpha}{\p
x^\gamma\p\tilde{x}^\beta}=0,
\]
so (\ref{sym0a}) indeed holds, i.e., \be\label{sym2} \tilde
c^\alpha_{\beta\gamma}=\tilde c^\alpha_{\gamma\beta}. \ee

Now we only need to prove that (\ref{ae2a}) holds,
or equivalently
\be\label{as1} \tilde c^\alpha_{\beta\gamma} \tilde
c^\gamma_{\rho\nu}-\tilde c^\alpha_{\nu\gamma} \tilde c^\gamma_{\rho\beta}=0. \ee
Using (\ref{sym2}),
(\ref{css2}) and (\ref{ccc}) we obtain \be\label{crep} \tilde
c^\alpha_{\beta\gamma}=\tilde c^\alpha_{\gamma\beta}=c^\alpha_{\gamma\varepsilon}\frac{\p x^\varepsilon}{\p \psi^\beta}=\frac{\p
x^\varepsilon}{\p \psi^\beta}c^\alpha_{\gamma\varepsilon}. \ee

Using (\ref{crep}) for $\tilde{c}^\alpha_{\beta\gamma}$ in the first
term of (\ref{as1}) and for $\tilde{c}^\alpha_{\nu\gamma}$ in the
second term of (\ref{as1}), and the formula that immediately follows
from (\ref{css2}),
\[
\tilde c^\varepsilon_{\kappa\rho}=c^\varepsilon_{\kappa\nu}\frac{\p x^\nu}{\p \psi^\rho},
\]
for $\tilde{c}^\gamma_{\rho\nu}$ in the first term of (\ref{as1})
and for $\tilde{c}^\gamma_{\rho\beta}$ in the second term of
(\ref{as1}), we see that the expression on the left-hand side of
(\ref{as1}) boils down to \be\label{as2a} \tilde
c^\alpha_{\beta\gamma} \tilde c^\gamma_{\rho\nu}-\tilde
c^\alpha_{\nu\gamma} \tilde c^\gamma_{\rho\beta} =\frac{\p x^\pi}{\p
\psi^\beta}\frac{\p x^\kappa}{\p
\psi^\nu}\left(c^\alpha_{\pi\gamma}c^\gamma_{\rho\kappa}-c^\alpha_{\kappa\gamma}
c^\gamma_{\rho\pi}\right)\stackrel{(\ref{ae0})}{=}0, \ee and thus
(\ref{ae2a}) indeed holds. $\square$

%\newpage
\section{Nonlocal symmetries for oriented associativity equations}
Recall (see e.g.\ \cite{olv_eng2,vin2,kk,blu})
that an (infinitesimal higher) symmetry for the oriented associativity equations (\ref{ae})
is an evolutionary vector field $X=G^\alpha\p/\p K^\alpha$ such that
$G^\alpha$ satisfy the linearized version of (\ref{ae}), that is,
\be\label{aesym}\ba{l} \ds\frac{\p^2 G^\nu}{\p x^\alpha\p
x^\rho}\frac{\p^2 K^\rho}{\p x^\beta\p x^\gamma}+\frac{\p^2 K^\nu}{\p x^\alpha\p
x^\rho}\frac{\p^2 G^\rho}{\p x^\beta\p x^\gamma} %\\[7mm]
\ds =\frac{\p^2 G^\rho}{\p x^\alpha\p x^\beta}
\frac{\p^2 K^\nu}{\p x^\rho\p x^\gamma}+\frac{\p^2 K^\rho}{\p x^\alpha\p x^\beta}
\frac{\p^2 G^\nu}{\p x^\rho\p x^\gamma},
\ea\ee
modulo (\ref{ae}) and differential consequences thereof (or, informally, on solutions of (\ref{ae})).
This is equivalent to compatibility of (\ref{ae}) with the flow associated with $X$, that is,
\[
\p K^\alpha/\p\tau=G^\alpha.
\]

A straightforward but somewhat tedious computation proves the following generalization
of the results of Chen, Kontsevich and Schwarz \cite{cks} (see Corollary \ref{sp1} below for the latter)
to the case of oriented associativity equations.
\begin{theo}\label{sp}%\vspace{-1mm}
%For any solution $\psi^\alpha(\lambda)$ of (\ref{zcr}) and any solution $\chi(\zeta)$
%of (\ref{zcrchi_oae})
The evolutionary vector fields \[
\psi^\alpha(\lambda)\ds\frac{\p}{\p K^\alpha}\quad\mbox{and}\quad
\psi^\alpha(\lambda)\chi(-\lambda)\ds\frac{\p}{\p K^\alpha},
\]
where $\psi^\alpha(\lambda)$ satisfy (\ref{zcr}) and $\chi(\lambda)$
satisfies (\ref{ssp_oae}), are nonlocal higher symmetries for the
oriented associativity equations (\ref{ae}), i.e., the flows
\begin{eqnarray}
\ds\frac{\p K^\alpha}{\p\tau_\lambda}=\psi^\alpha(\lambda),\label{nls0}\\
%\ee
%where $\psi^\alpha$ satisfy (\ref{zcr}), is compatible %(``commutes")
%with (\ref{ae}), and the system
%\be
\label{nls1}
\ds\frac{\p K^\alpha}{\p\sigma_\lambda}=\psi^\alpha(\lambda)\chi(-\lambda),
\end{eqnarray}
are compatible %(``commute")
with (\ref{ae}).
\end{theo}
Informally, compatibility here means that the flows (\ref{nls0})  and (\ref{nls1})
map the set $\mathcal{S}$ of (smooth) solutions of (\ref{ae}) into itself, i.e.,
$\mathcal{S}$ is invariant under the flows (\ref{nls0}) and (\ref{nls1}); see e.g.\ \cite{vin2,kk,blu,kon_mok,
vin0,vin1} and references therein
for the general theory of nonlocal symmetries. In a more analytic language,
Theorem~\ref{sp} states that $G^\alpha=\psi^\alpha(\lambda)$ and $\tilde G^\alpha=\psi^\alpha(\lambda)\chi(-\lambda)$
satisfy (\ref{aesym}) provided (\ref{ae}), (\ref{zcr}) and (\ref{ssp_oae}) hold.\looseness=-1

An unusual feature of the symmetries $\psi^\alpha(\lambda)\p/\p K^\alpha$ from Theorem~\ref{sp} is that
they are linear (rather than quadratic, as it
is the case for many other systems, cf.\ \cite{konbook}) in the solutions of auxiliary linear problem.
\looseness=-1

It is natural to ask whether the flows (\ref{nls0}) and (\ref{nls1}) are integrable systems
in any reasonable sense. The following result provides
linear spectral problems for these flows and thus suggests their integrability.\looseness=-1
\begin{cor}\label{extfl}
The flows (\ref{nls0}) and (\ref{nls1}) can be (nonuniquely)
extended to the flows for the quantities $\psi^\alpha(\mu)$ and $\chi(\mu)$ as follows:
\begin{eqnarray}
\ds\frac{\p \psi^\alpha(\mu)}{\p\tau_\lambda}=\lambda\mu c^\alpha_{\nu\kappa}\psi^\nu(\lambda)\psi^\kappa(\mu),\label{nls0psi}\\
\ds\frac{\p \chi(\mu)}{\p\tau_\lambda}=\frac{\lambda\mu}{\lambda+\mu}
\psi^\nu(\lambda)\frac{\p\chi(\mu)}{\p x^\nu},\label{nls0chi}\\
\ds\frac{\p \psi^\alpha(\mu)}{\p\sigma_\lambda}=\lambda\mu c^\alpha_{\nu\kappa}\psi^\nu(\lambda)\psi^\kappa(\mu)\chi(-\lambda)
+\frac{\lambda\mu}{\lambda-\mu}\frac{\p\chi(-\lambda)}{\p x^\beta}\psi^\beta(\mu)\psi^\alpha(\lambda),\label{nls1psi}\\
\ds\frac{\p \chi(\mu)}{\p\sigma_\lambda}=\frac{\lambda\mu}{\lambda+\mu} \psi^\nu(\lambda)\frac{\p\chi(\mu)}{\p x^\nu}
\chi(-\lambda).\label{nls1chi}
\end{eqnarray}
\end{cor}
%{\bf Remark 1} Note some important implications of
%Corollary~\ref{extfl}:
%\begin{itemize}
%\item
In particular, by Corollary~\ref{extfl} Eq.(\ref{nls0psi}) together with the system
\be\label{zcrmu}
\frac{\p \psi^\alpha(\mu)}{\p x^\beta}=\mu \frac{\p^2 K^\alpha}{\p x^\beta\p x^\gamma}\psi^\gamma(\mu),
\ee
provide (assuming that (\ref{zcr}) holds) a zero-curvature representation for the extended system (\ref{ae}), (\ref{nls0}),
%i.e.,
%this extended system is precisely the set of compatibility conditions for (\ref{nls0psi}) and (\ref{zcrmu}),
and thus ensure integrability thereof.\looseness=-1

Likewise, the flow (\ref{nls1}) is integrable because Eq.(\ref{zcrmu}) along with the system
\be\label{zcrchimu}
\frac{\p^2 \chi(\mu)}{\p x^\alpha\p x^\beta}=\mu \frac{\p^2 K^\delta}{\p x^\alpha\p x^\beta}\frac{\p\chi(\mu)}{\p x^\delta}
\ee
and (\ref{nls1psi}), (\ref{nls1chi})
provide (assuming that (\ref{zcr}) and (\ref{ssp_oae}) hold) a linear spectral problem for the extended system (\ref{ae}), (\ref{nls1}), i.e., (\ref{ae}) and (\ref{nls1})
are precisely the compatibility conditions for (\ref{nls1psi})--(\ref{zcrchimu}), and integrability of
the extended system in question follows.\looseness=-1
%\end{itemize}

%Using Corollary~\ref{extfl} we can readily find the commutators of the nonlocal
%symmetries of (\ref{ae}),i.e., of nonlocal evolutionary vector fields
%\be\label{nlevf}
%X(\lambda)=\psi^\alpha(\lambda)\frac{\p}{\p K^\alpha},\quad Y(\lambda)=\psi^\alpha(\lambda)\chi(-\lambda)\frac{\p}{\p K^\alpha}.
%\ee
Using the extended flows from Corollary~\ref{extfl} we readily obtain the following result:
\begin{cor}\label{extfl1} All flows (\ref{nls0}) and (\ref{nls1})
commute for all values of parameters $\lambda$ and $\mu$:
\begin{eqnarray*}\label{comrel}
%\hspace*{-15mm}
\ds\frac{\p^2 K^\alpha}{\p\tau_\lambda\p\tau_{\mu}}\!=\!\frac{\p^2 K^\alpha}{\p\tau_{\mu}\p\tau_{\lambda}},\quad %\\ %[12mm]
\ds\frac{\p^2 K^\alpha}{\p\tau_{\lambda}\p\sigma_{\mu}}\!=\!\frac{\p^2 K^\alpha}{\p\sigma_{\mu}\p\tau_{\lambda}},\quad %\\ %[13mm]
\ds\frac{\p^2 K^\alpha}{\p\sigma_{\lambda}\p\sigma_{\mu}}\!=\!\frac{\p^2 K^\alpha}{\p\sigma_{\mu}\p\sigma_{\lambda}}.
%, k,l=0,1,2,\dots,\ \alpha,\beta,\gamma,\nu,q=1,\dots,n.
%\ea
%\qquad\\ba{l}
%k,l=0,1,2,\dots,\quad \alpha,\beta,\gamma,\delta,\nu,\nu=1,\dots,n.\nonumber
\end{eqnarray*}
\end{cor}

It is important to stress that this result {\em per se} does {\em
not} imply commutativity of the {\em extended} flows from
Corollary~\ref{extfl}. However, a straightforward computation proves
the following assertion.
\begin{cor}\label{extfl2}The extended flows (\ref{nls0}), (\ref{nls0psi}), (\ref{nls0chi}) commute, i.e.,
for all values of $\lambda$, $\mu$ and $\zeta$
we have
\be\label{ext_fl_c}
\ds\frac{\p^2 K^\alpha}{\p\tau_\lambda\p\tau_{\mu}}\!=\!\frac{\p^2 K^\alpha}{\p\tau_{\mu}\p\tau_{\lambda}},\quad
\ds\frac{\p^2 \psi^\alpha(\zeta)}{\p\tau_\lambda\p\tau_{\mu}}\!=\!\frac{\p^2 \psi^\alpha(\zeta)}{\p\tau_{\mu}\p\tau_{\lambda}},\quad
\ds\frac{\p^2  \chi(\zeta)}{\p\tau_\lambda\p\tau_{\mu}}\!=\!\frac{\p^2 \chi(\zeta)}{\p\tau_{\mu}\p\tau_{\lambda}}.
\ee
\end{cor}
We intend to study the remaining commutation relations
for the extended flows in more detail elsewhere.

\section{Expansion in the spectral parameter and nonlocal potentials}
Now consider a formal Taylor expansion for $\psi^\alpha$ in
$\lambda$, %\vspace{-3mm}
\be\label{psifor}
\psi^\alpha(\lambda)=\sum\limits_{k=0}^\infty
\psi^\alpha_k\lambda^k.%\vspace{-2mm}
\ee
It is immediate from Theorem~\ref{sp} that $\psi_k^\alpha\p/\p
K^\alpha$ are symmetries for (\ref{ae}), i.e., the flows %\vspace{-1mm}
\be\label{psik0}%\vspace{-5mm}
\frac{\p K^\alpha}{\p\tau_k}=\psi_k^\alpha,\quad k=0,1,2,\dots,
\ee
are compatible with (\ref{ae}).

We readily find from (\ref{zcr}) the following recursion relation:
%for symmetries of (\ref{ae}):
\be\label{recc}
  \frac{\p\psi^\alpha_{k}}{\partial x^{\beta}}=
\frac{\p^2 K^\alpha}{\p x^\beta\p x^\gamma}
  \psi^\gamma_{k-1},\quad k=1,2,\dots.
\ee

For $k=0$ we have $\partial\psi^\alpha_{0}/\partial x^{\beta}=0$
%\be\label{recc0}
% $\frac{\partial\psi^\alpha_{0}}{\partial x^{\beta}}=0$
%\ee
for all $\beta=1,\dots,n$.

Now let $(w_0)^\alpha_\beta=\delta^\alpha_\beta$, where $\delta^\alpha_\beta$ is the Kronecker delta,
and $(w_1)^\alpha_\beta=\p K^\alpha/\p x^\beta$. Define recursively
the following sequence of nonlocal quantities:
\be\label{wrec}
\frac{\partial (w_k)_\gamma^\beta}{\partial x^{\alpha}}=
\frac{\p^2 K^\beta}{\p x^\alpha\p
x^\rho}(w_{k-1})^\rho_\gamma,\quad k=2,3,\dots
\ee

We have %obtain the following generalization of (\ref{psi2}):
\be\label{psik} \psi_k^\alpha=\sum\limits_{j=0}^{k}h_{j}^\gamma
(w_{k-j})^\alpha_\gamma, \quad k=0,1,2,\dots,
\ee
where $h_j^\gamma$
are arbitrary constants.

%As a linear combination of symmetries is again a symmetry, Theorem~\ref{sp} along with (\ref{psifor})
%implies that any $\psi_k=(\psi_k^1,\dots,\psi_k^n)$ is a
%symmetry for (\ref{ae}). Moreover, using (\ref{psik}) we readily obtain the following result.

%\newpage

In analogy with (\ref{psifor}), %we can make
consider
a formal Taylor expansion for $\chi$ in $\lambda$, %\vspace{-3mm}
\be\label{chiexp}
\chi(\lambda)=\sum\limits_{k=0}^\infty \chi_k\lambda^k.
%\vspace{-3mm}
\ee

We obtain from (\ref{ssp_oae}) the following recursion relation:
%for symmetries of (\ref{wdvv}):
\be\label{rec_chi_oae}
  \frac{\partial^2\chi_k}{\partial x^{\alpha} \partial x^{\gamma}}=
\frac{\partial^2 K^\nu}{\partial x^{\alpha} \partial x^{\gamma}}
 \frac{\partial\chi_{k-1}}{\partial x^{\nu}},\quad k=1,2,\dots.
\ee

%For $k=0$ we have
%\be\label{rec0_oae}
%  \frac{\partial^2\chi_0}{\partial x^{\alpha} \partial x^{\gamma}}=0.
%\ee
%whence
%\be\label{ic00_oae}\chi_0=b_0+d_{0,\gamma} x^\gamma,\nopagebreak[4]
%\ee{}
%with $b_0$ and $d_{0,\gamma}$ being arbitrary constants.
%%
%Plugging (\ref{ic00_oae}) into (\ref{rec_chi_oae}) for $k=1$ yields
%\be\label{chi1ord}
%\chi_1=b_1+d_{1,\gamma} x^\gamma+ d_{0,\beta}K^{\beta},
%\ee
%where $d_{1,\gamma}$ and $b_1$ again are arbitrary constants.
%\looseness=-1

Set $v_0^\alpha=x^\alpha$, $v_1^\alpha=K^\alpha$, and, in analogy with (\ref{chiaexp}) and (\ref{wrec}),
define the following sequence of nonlocal quantities
$v^\beta_k$, $\beta=1,\dots,n$: %, $k=0,1,2,\dots$:
\begin{eqnarray}\label{vdef_oae}
%\frac{\partial^2 v_{0}^\beta}{\partial x^{\alpha}\partial x^{\gamma}}=
%\frac{\p^2 K^\nu}{\p x^\alpha\p x^\gamma}\frac{\p K^\beta}{\p x^\nu},\\
\frac{\partial^2 v_{k}^\beta}{\partial x^{\alpha}\partial x^{\gamma}}=
\frac{\p^2 K^\nu}{\p x^\alpha\p x^\gamma}\frac{\p v_{k-1}^\beta}{\p x^\nu},\quad k=2,3,\dots \label{vrec_oae}
\end{eqnarray}
In terms of
geometric theory of PDEs, see e.g.\ \cite{vin2,kk,vin0,vin1}, the
quantities $(w_k)_\beta^\alpha$ and $v_k^\gamma$, $\alpha,\beta,\gamma=1,\dots,n$,
$k=2,3,\dots$, define an
infinite-dimensional Abelian covering over (\ref{ae}).

We have
the following counterpart of (\ref{psik}):%\vspace{-5mm}
\be\label{chik_oae}
\chi_k=b_k+\sum\limits_{j=0}^{k}d_{k-j,\gamma} v_{j}^\gamma,\quad k=0,1,2,\dots,
\ee
where $b_k$ and $d_{j,\gamma}$ %and $h_{j}^\gamma$
are arbitrary constants.

Using (\ref{psifor}) and (\ref{chiexp}) we readily find that
\[
\psi^\alpha(\lambda)\chi(-\lambda)=\sum\limits_{k=0}^\infty\rho^\alpha_k\lambda^k,\quad
\rho^\alpha_k\stackrel{\mathrm{def}}{=}\sum\limits_{j=0}^k(-1)^j\chi_j \psi^\alpha_{k-j}.
\]
%where
%\[
%\rho^\alpha_k=\sum\limits_{j=0}^k(-1)^j\chi_j \psi^\alpha_{k-j}.
%\]
It is now immediate from Theorem~\ref{sp} that $(w_k)^\alpha_\beta\p/\p K^\alpha$ and
$\rho_k^\alpha\p/\p K^\alpha$ are symmetries for (\ref{ae}),
and for $k\geq 2$ these symmetries are nonlocal.

What is more, using Corollary~\ref{extfl} we readily obtain the following result.
%find that all these lifted flows commute:

%\newpage
\begin{cor}\label{oae_new_sym2}
The oriented associativity equations (\ref{ae}) have infinitely many symmetries of the form
\[
X_{k,\beta}=(w_k)_\beta^\alpha\frac{\p}{\p
K^\alpha}\quad\mbox{and}\quad
Y_{k,\gamma}^\beta=\quad\ds\!\!\ds\sum\limits_{j=0}^k(-1)^j
v_j^\beta\cdot (w_{k-j})^\alpha_\gamma\frac{\p}{\p K^\alpha},
\]
and all associated flows, i.e.,
\[
\ds \frac{\p K^\alpha}{\p\tau^\beta_{k}}=(w_k)_\beta^\alpha,\ \
\ds \frac{\p K^\alpha}{\p\sigma^\gamma_{k,\beta}}
=\!\!\ds\sum\limits_{j=0}^k(-1)^j v_j^\beta\cdot (w_{k-j})^\alpha_\gamma,
\]
%lifted to the space of variables $(w_k)_\beta^\gamma$ and $v_k^\beta$
%constructed using Corollary~\ref{extfl},
commute:
\[
\ba{l}
%\hspace*{-15mm}
\ds\frac{\p^2 K^\alpha}{\p\tau^\beta_{k}\p\tau^\nu_{l}}\!=\!\frac{\p^2 K^\alpha}{\p\tau^\nu_{l}\p\tau^\beta_{k}},\qquad
\ds\frac{\p^2 K^\alpha}{\p\sigma^\beta_{k,\gamma}\p\tau^\rho_{l,\nu}}
\!=\!\frac{\p^2 K^\alpha}{\p\tau^\rho_{l,\nu}\p\sigma^\beta_{k,\gamma}}, \qquad %\\ %[13mm]
\ds\frac{\p^2 K^\alpha}{\p\sigma^\beta_{k,\gamma}\p\sigma^\rho_{l,\nu}}\!=\!\frac{\p^2 K^\alpha}{\p\sigma^\rho_{l,\nu}
\p\sigma^\beta_{k,\gamma}},\\[7mm] %[14mm]
%, k,l=0,1,2,\dots,\ \alpha,\beta,\gamma,\pi,q=1,\dots,n.
%\ea
%\qquad\\ba{l}
k,l=0,1,2,\dots,\quad \alpha,\beta,\gamma,\delta,\nu,\rho=1,\dots,n.
\ea
\]
\end{cor}

It is readily seen that the symmetries $X_{k,\beta}$ and $Y_{k,\gamma}^\beta$
and the associated flows are nonlocal for $k\geq 2$. We stress once more that
the part of Corollary~\ref{oae_new_sym2}
about commutativity of the flows associated with
$X_{k,\beta}$ and $Y_{k,\gamma}^\beta$ makes substantial use of the extended flows from Corollary~\ref{extfl}
which are not uniquely defined. Moreover, Corollary~\ref{oae_new_sym2} does {\em not}
imply commutativity of the {\em extended} flows associated with the symmetries
$X_{k,\beta}$ and $Y_{k,\gamma}^\beta$. However, the flows associated with the symmetries
$X_{k,\beta}$ do commute after a (suitable) extension to the variables $(w_k)^\alpha_\beta$.
Namely, using Corollaries~\ref{extfl}~and~\ref{extfl2}
we readily arrive at the following assertion.\looseness=-1
\begin{cor}\label{extfl_x}
The flows
\be\label{wkflow}\ba{c}
%\mbox{Let}\
\ds \frac{\p K^\alpha}{\p\tau^\beta_k}=(w_k)_\beta^\alpha,\\[5mm]
\ds \frac{\p (w_l)_\pi^\alpha}{\p\tau^\beta_k}=
c^\alpha_{\nu\rho}(w_{k-1})_\beta^\nu (w_{l-1})_\pi^\rho, \quad l=0,1,2,\dots,\ea %\quad k=0,1,2,\dots,
\ee
where $(w_0)^\alpha_\beta=\delta^\alpha_\beta$, $(w_1)^\alpha_\beta=\p K^\alpha/\p x^\beta$,
and we set $(w_{-1})_\beta^\nu\equiv 0$ for convenience, {\em commute}
for all $\beta,\gamma=1,\dots,n$ and all $k,l=0,1,2,\dots$.

Equivalently, the
evolutionary vector fields
\[
\bar{X}_{s,\beta}\equiv X_{s,\beta}=(w_s)_\beta^\alpha\frac{\p}{\p K^\alpha},\quad
s=0,\quad \bar{X}_{k,\beta}=(w_k)_\beta^\alpha\frac{\p}{\p
K^\alpha}+\sum\limits_{l=2}^{\infty}c^\alpha_{\kappa\rho}(w_{k-1})_\beta^\kappa
(w_{l-1})_\pi^\rho\frac{\p}{\p (w_l)_\pi^\alpha}, \quad k=1,2,3,\dots,
\]
{\em commute}, i.e., $[\bar{X}_{k,\beta},\bar{X}_{l,\gamma}]=0$ (see e.g.\ \cite{vin2} for the definition of the bracket $[,]$),
%\be\label{cvf1}
%[X_{\beta,k},X_{\gamma,l}]=0
%\ee
for all $\beta,\gamma=1,\dots,n$ and all $k,l=0,1,2,\dots$.
\end{cor}
Thus, the oriented associativity equations (\ref{ae}) possess an infinite hierarchy of commuting flows
whose existence reconfirms integrability of (\ref{ae}).

\section{Nonlocal symmetries for the gradient reduction\\ of oriented associativity equations}
Following \cite{kon}, consider the so-called gradient reduction of (\ref{ae}). Namely, assume that there exist
a nondegenerate symmetric constant matrix $\eta^{\alpha\beta}$ and a function $F=F(x^1,\dots,x^n)$,
known as a prepotential in 2D topological field theories \cite{wi,dvv,dub},
such that
\be\label{gr}
K^\alpha=\eta^{\alpha\beta}\p F/\p x^\beta.
\ee

%Following \cite{dub} consider the WDVV equations for an unknown
%function $F=F(\vec x)$, where $\vec x=(x^1,\dots,x^n)$. These equations read
Then (\ref{ae}) boils down to the famous associativity equations for $F$ \cite{wi,dvv,dub, dub2}:
\begin{align}
\label{wdvv}
  %\sum_{\delta,c=1}^n
  \frac{\partial^3 F}{\partial x^{\alpha}
   \partial x^{\beta} \partial x^{\delta} } \eta^{\delta\gamma}
   \frac{\partial^3 F}{\partial x^{\gamma} \partial x^{\nu}
   \partial x^{\rho}}
   =%\sum_{\delta, c=1}^n
  \frac{\partial^3 F}{\partial x^{\alpha} \partial
   x^{\nu} \partial x^{\delta} }
  \eta^{\delta\gamma} \frac{\partial^3 F}{\partial x^\gamma
   \partial x^\beta \partial x^\rho}.
\end{align}
%where $\eta^{\alpha\beta}$ is a nondegenerate symmetric constant
%matrix.

Note that in the standard theory of the WDVV equations (see e.g.\  \cite{wi, dvv, dub, dub2}) it is further required that
\begin{align}
\label{etac}
  \frac{\partial^3 F}{\partial x^\alpha \partial x^\beta \partial x^1 }
  = \eta_{\alpha\beta},
\end{align}
where $\eta_{\alpha\beta}$ is a nondegenerate constant matrix
such that $\eta_{\alpha\beta}\eta^{\beta\gamma}=\delta_\alpha^\gamma$.
%\[
%\eta_{\alpha\beta}\eta^{\beta\gamma}=\delta_\alpha^\gamma.
%\]

However, in what follows we shall not impose (\ref{etac}) and
the so-called quasihomogeneity condition (see e.g.\ \cite{wi, dvv, dub,dub2,manin_bk}
for the discussion of these conditions).
%Here and below we assume the Einstein summation convention, with the
%sum over any pair of repeated indices running from 1 to $n$ unless
%otherwise explicitly stated.

Upon assuming (\ref{gr}) we find that the auxiliary linear problem (\ref{zcr})
also admits a reduction
\be\label{grzcr}
\psi^\alpha=\eta^{\alpha\beta}\p\chi/\p x^\beta.
\ee
This, along with (\ref{gr}), turns (\ref{zcr}) into the following overdetermined
system of the Gauss--Manin equations for $\chi$:\looseness=-1
\begin{equation}
\label{zcrchi}
  \frac{\partial^2\chi(\lambda)}{\partial x^{\alpha} \partial x^{\gamma}}= \lambda
\eta^{\rho\nu} \frac{\partial^3 F}{\partial x^{\alpha} \partial x^{\gamma}\partial x^{\rho}}
 \frac{\partial\chi(\lambda)}{\partial x^{\nu}}.
\end{equation}
This is nothing but (\ref{ssp_oae}) after the substitution (\ref{gr}), and again
the associativity equations (\ref{wdvv}) are nothing but
the compatibility conditions for (\ref{zcrchi});
see e.g.\ \cite{dub,dub2,fer} for the discussion of geometric aspects
of (\ref{zcrchi}).
%, and \cite{kon} for the interpretation of $\chi$ as a wave function
%in the context of quantum deformations of associative algebras.\looseness=-1

%We shall first consider (\ref{wdvv}) without imposing (\ref{etac}).
%Then a straightforward but tedious computation yields the following
Using Theorem~\ref{sp} in conjunction with (\ref{gr}) and (\ref{grzcr}) we recover the following
result from
%readily obtain the following result,
%which is essentially equivalent to that of
\cite{cks}.
\begin{cor}\label{sp1}
%\vspace{-2mm}
For any solution $\chi(\lambda)$ of (\ref{zcrchi})
the quantities
\[
\chi(\lambda)\ds\frac{\p}{\p F}\quad\mbox{and}\quad\chi(\lambda)\chi(-\lambda)\ds\frac{\p}{\p F}
\]
are nonlocal higher symmetries
for the associativity equations (\ref{wdvv}), i.e., the equations
\begin{eqnarray}
\frac{\p F}{\p\tau_\lambda}=\chi(\lambda),\label{nls0af}\\
\label{nls1af}
\frac{\p F}{\p\zeta_\lambda}=\chi(\lambda)\chi(-\lambda)
\end{eqnarray}
%where $\chi(\lambda)$ satisfies (\ref{zcrchi}),
are compatible with
(\ref{wdvv}).%\vspace{-2mm}

%Likewise, $\chi(\lambda)\chi(-\lambda)$ also is a nonlocal higher symmetry
%for the associativity equations (\ref{wdvv}), i.e., the equation
%\be\label{nls1af}
%\frac{\p F}{\p\zeta_\lambda}=\chi(\lambda)\chi(-\lambda),
%\ee
%where $\chi(\lambda)$ satisfies (\ref{zcrchi}), is compatible with
%(\ref{wdvv}).%\vspace{-2mm}

The above flows can be (nonuniquely) extended as follows
\begin{eqnarray}
\ds\frac{\p \chi(\mu)}{\p\tau_\lambda}=\frac{\lambda\mu}{\lambda+\mu}
\eta^{\nu\beta}\frac{\p\chi(\lambda)}{\p x^\beta}\frac{\p\chi(\mu)}{\p x^\nu},\label{nlswdvv0chi}\\
\ds\frac{\p \chi(\mu)}{\p\zeta_\lambda}=\frac{\lambda\mu}{\lambda+\mu}
\eta^{\nu\beta}\frac{\p\chi(\lambda)}{\p x^\beta}\frac{\p\chi(\mu)}{\p x^\nu}\chi(-\lambda)
+\frac{\lambda\mu}{\lambda-\mu} \eta^{\nu\beta}\frac{\p\chi(-\lambda)}
{\p x^\beta}\frac{\p\chi(\mu)}{\p x^\nu}\chi(\lambda).\label{nlswdvv1chi}
\end{eqnarray}
\end{cor}
In particular, this result means that
$\chi(\lambda)$ and $\chi(\lambda)\chi(-\lambda)$ satisfy
the linearized version of (\ref{wdvv}) provided
(\ref{wdvv}) and (\ref{zcrchi}) hold.
Using the extended flows from Corollary~\ref{sp1} we readily obtain
%the following result.
\begin{cor} All flows (\ref{nls0af}) and (\ref{nls1af})
commute: for all values of parameters $\lambda$ and $\mu$ we have
\begin{eqnarray*}\label{comrel_wdvv}
%\hspace*{-15mm}
\ds\frac{\p^2 F}{\p\tau_\lambda\p\tau_{\mu}}\!=\!\frac{\p^2 F}{\p\tau_{\mu}\p\tau_{\lambda}},\quad %\\ %[12mm]
\ds\frac{\p^2 F}{\p\tau_{\lambda}\p\zeta_{\mu}}\!=\!\frac{\p^2 F}{\p\zeta_{\mu}\p\tau_{\lambda}},\quad %\\ %[13mm]
\ds\frac{\p^2 F}{\p\zeta_{\lambda}\p\zeta_{\mu}}\!=\!\frac{\p^2 F}{\p\zeta_{\mu}\p\zeta_{\lambda}}.
%, k,l=0,1,2,\dots,\ \alpha,\beta,\gamma,\nu,q=1,\dots,n.
%\ea
%\qquad\\ba{l}
%k,l=0,1,2,\dots,\quad \alpha,\beta,\gamma,\delta,\nu,\nu=1,\dots,n.\nonumber
\end{eqnarray*}
\end{cor}

Perhaps a bit surprisingly, the proper counterpart of the flow (\ref{nls1af}) for the oriented associativity equations (\ref{ae})
is not (\ref{nls1}) itself but a linear combination of the flows (\ref{nls1}) with the opposite values of $\lambda$:
\[
\frac{\p K^\alpha}{\p\zeta_\lambda}=\psi^\alpha(\lambda)\chi(-\lambda)+\psi^\alpha(-\lambda)\chi(\lambda).
\]

%Also note that it readily follows from the results of \cite{cks} that ceratin superpositions of symmetr (\ref{nls1af})
%is compatible with the condition (\ref{etac}).

%Just as for Theorem~\ref{rp},

Consider now
a formal Taylor expansion  in $\lambda$ for a solution $\chi(\lambda)$ of (\ref{zcrchi}),%\vspace{-3mm}
\[
\chi(\lambda)=\sum\limits_{k=0}^\infty \chi_k\lambda^k. %\vspace{-3mm}
\]
Notice that using
a slightly different expansion of $\chi(\lambda)$,
involving also $\lambda^{-1}$, enables one to construct solutions
of the WDVV equations directly from $\chi(\lambda)$, see \cite{avdl} and references therein.

%The formulas
Eqs.(\ref{chik_oae}) remain valid
when $\chi(\lambda)$ satisfies (\ref{zcrchi}) instead of (\ref{ssp_oae})
if we substitute $\eta^{\alpha\beta}\p F/\p x^\beta$ for $K^\alpha$ into the definitions
of $v_k^\alpha$ and (\ref{chik_oae}). Then %With this in mind,
expanding the symmetries from Corollary~\ref{sp1} in powers of $\lambda$ yields
\begin{cor}\label{oae_new_sym2wdvv}
The associativity equations (\ref{wdvv}) have infinitely many symmetries of the form
\[
\tilde X^\beta_k=v_k^\beta\frac{\p}{\p F},\quad\mbox{and}
\quad\tilde Z^{\alpha\beta}_k=\ds\sum\limits_{j=0}^k(-1)^j
v_j^\alpha v_{k-j}^\beta\frac{\p}{\p F},\quad k=0,1,2,\dots,\quad \alpha,\beta=1,\dots,n,
\]
and all associated flows, i.e.,
\be\label{wdvvsymfl}
\ds \frac{\p F}{\p\tau_{\beta,k}}=v_k^\beta,\ \
\ds \frac{\p F}{\p\zeta_{\alpha\beta,k}}=\!\!\ds\sum\limits_{j=0}^k(-1)^j v_j^\alpha v_{k-j}^\beta,
\ee
%lifted to the space of variables $(w_k)_\beta^\gamma$ and $v_k^\beta$
%constructed using Corollary~\ref{extfl},
commute:
\[
\ba{l}
%\hspace*{-15mm}
\ds\frac{\p^2 F}{\p\tau_{\beta,k}\p\tau_{\nu,l}}\!=\!\frac{\p^2 F}{\p\tau_{\nu,l}\p\tau_{\beta,k}},\qquad
\ds\frac{\p^2 F}{\p\tau_{\gamma,k}\p\zeta_{\alpha\beta,l}}\!=\!\frac{\p^2 F}{\p\zeta_{\alpha\beta,l}\p\tau_{\gamma,k}},\qquad
\ds\frac{\p^2 F}{\p\zeta_{\rho\nu,k}\p\zeta_{\alpha\beta,l}}\!=\!\frac{\p^2 F}{\p\zeta_{\alpha\beta,l}\p\zeta_{\rho\nu,k}},\\[7mm]
%, k,l=0,1,2,\dots,\ \alpha,\beta,\gamma,\nu,q=1,\dots,n.
%\ea
%\qquad\\ba{l}
k,l=0,1,2,\dots,\quad \alpha,\beta,\gamma,\delta,\nu,\rho=1,\dots,n.
\ea
\]
\end{cor}

Again, it is clear that the symmetries $\tilde X^\beta_k$ and $\tilde Z^{\alpha\beta}_k$ and the associated flows
are nonlocal for $k\geq 2$.

Note that the above results do {\em not} imply commutativity for {\em all} of
the {\em extended} flows from Corollary~\ref{sp1} and therefore do not contradict
the results from \cite{giv, cks}. On the other hand, using Corollary~\ref{extfl2}
we readily find that, in perfect agreement with \cite{cks}, the extended flows (\ref{nls0af}), (\ref{nlswdvv0chi})
do commute, i.e.,
\be\label{ext_fl_comm_wdvv}
\ds\frac{\p^2 F}{\p\tau_\lambda\p\tau_{\lambda'}}\!=\!\frac{\p^2 F}{\p\tau_{\lambda'}\p\tau_{\lambda}},\quad
\ds\frac{\p^2 \chi(\mu)}{\p\tau_\lambda\p\tau_{\lambda'}}\!=\!\frac{\p^2 \chi(\mu)}{\p\tau_{\lambda'}\p\tau_{\lambda}},
\ee
for all values of $\lambda$, $\lambda'$ and $\mu$.

The quantities $\chi_k$ coincide, up to a choice of normalization,
with the densities of Hamiltonians of
integrable bihamiltonian hydrodynamic-type systems associated to
any solution of the WDVV equations, see Lecture 6 of \cite{dub2} and e.g.\ \cite{chenetc}, and references therein.
It was mentioned in \cite{dub2} that it is natural to consider these hydrodynamic-type
systems as higher (Lie--B\"acklund)
symmetries for the WDVV equations,
because using these systems one can
construct \cite{dub2} the B\"acklund transformation for the WDVV equations.
We have now seen that $\chi_k$ and $v_k^\alpha$ can also be interpreted
as symmetries of the associativity equations (\ref{wdvv})
(and, upon imposing necessary restrictions on $F$ and $\chi$, of the WDVV equations)
in a far more straightforward manner.\looseness=-1

\section{Intermediate integrals and B\"acklund-type transformations}
The compatibility conditions
\[
\frac{\partial^2 (w_2)_\gamma^\beta}{\partial x^{\alpha}\partial
x^\nu} =\frac{\partial^2 (w_2)_\gamma^\beta}{\partial
x^\nu\partial x^{\alpha}}
\]
for (\ref{wrec}) with $k=2$ yield precisely Eqs.(\ref{ae}), and we have
\begin{cor}\label{ii1}
If the functions $K^\alpha$ and
$G^\beta_\gamma$
%$\alpha,\beta,\gamma=1,\dots,n$,
satisfy the system
\be\label{firint}%\vspace{-5mm}
\frac{\p^2 K^\beta}{\p x^\alpha\p x^\rho}\frac{\p K^\rho}{\p x^\gamma}=
\ds\frac{\p G^\beta_\gamma}{\p x^\alpha},\ %\alpha,\beta,\gamma=1,\dots,n,
\ee
then $K^\alpha$ automatically satisfy the oriented associativity equations (\ref{ae}).

In particular, if under the above assumptions
$K^\alpha=\eta^{\alpha\beta}\p F/\p x^\beta$, where $\eta^{\alpha\beta}$ is
a symmetric nondegenerate constant matrix,
then $F$ automatically satisfies the associativity equations (\ref{wdvv}).
\end{cor}

Just like (\ref{firint}), the compatibility conditions
\[
\frac{\partial^3 v_2^\nu}{\partial x^{\alpha}\partial
x^\beta\partial x^\gamma} =\frac{\partial^3 v_2^\nu}{
\partial x^\gamma \partial x^\beta \partial x^{\alpha}}
\]
for (\ref{vrec_oae}) with $k=2$ also yield nothing but
Eqs.(\ref{ae}), and we obtain
\begin{cor}\label{ii2}
If the functions $K^\alpha$ and $G^\beta$ satisfy
\be\label{firint2}%\vspace{-5mm}
\ds\frac{\p^2 K^\nu}{\p x^\alpha\p x^\gamma}\frac{\p K^\beta}{\p
x^\nu}= \frac{\partial^2 G^\beta}{\partial x^{\alpha}\partial
x^{\gamma}},\ %\alpha,\beta,\gamma=1,\dots,n,
\ee
then $K^\alpha$ automatically satisfy the oriented associativity equations (\ref{ae}).

In particular, if under the above assumptions
$K^\alpha=\eta^{\alpha\beta}\p F/\p x^\beta$, where $\eta^{\alpha\beta}$ is
a symmetric nondegenerate constant matrix,
then $F$ automatically satisfies the associativity equations (\ref{wdvv}).
\end{cor}

Thus, (\ref{firint}) and (\ref{firint2}) yield a kind of intermediate integrals
for (\ref{ae}) (or, if $K^\alpha=\eta^{\alpha\beta}\p F/\p x^\beta$, for (\ref{wdvv})).
In the terminology of \cite{vin2,kk,vin0,vin1}, (\ref{firint}) and (\ref{firint2}) also define coverings
over (\ref{ae}) (respectively, if $K^\alpha=\eta^{\alpha\beta}\p F/\p x^\beta$, over (\ref{wdvv})):
for any solution $K^\alpha$ of (\ref{ae}) (resp.\
for any solution $F$ of (\ref{wdvv})) there exist,
at least locally, the functions $G^\alpha_\beta$ and
$G^\alpha$ such that (\ref{firint}) and (\ref{firint2}) hold.\looseness=-1

Moreover, we have
the following observation.
%This implies the following easy observation:
\begin{cor}\label{bt}
If the functions $K^\alpha$ %=\check{K}^\alpha(x^1,\dots,x^n)$ and
and $H^\alpha$ %=\check{H}^\alpha(x^1,\dots,x^n)$
satisfy the system %\vspace{-5mm}
\be\label{firint_oae} \frac{\p^2 K^\beta}{\p x^\alpha\p
x^\rho}\frac{\p K^\rho}{\p x^\gamma}
=\ds\frac{\p^2 H^\beta}{\p x^\alpha\p x^\gamma}, %\alpha,\beta,\gamma=1,\dots,n,
\ee
then $K^\alpha$ and $\tilde K^\alpha=H^\alpha$ automatically satisfy the oriented associativity equations (\ref{ae}).
\end{cor}

Hence, (\ref{firint_oae}) provides a conditional B\"acklund transformation
for (\ref{ae}): if $K^\alpha$ satisfy (\ref{ae}) {\em and} the conditions
\be\label{sym_cond}
\frac{\p^2 K^\beta}{\p x^\alpha\p
x^\rho}\frac{\p K^\rho}{\p x^\gamma}
=\frac{\p^2 K^\beta}{\p x^\gamma\p
x^\rho}\frac{\p K^\rho}{\p x^\alpha}
\ee
which are necessary for (\ref{firint_oae}) to hold,
then there exist, at least locally, $\tilde K^\alpha=H^\alpha$
such that (\ref{firint_oae}) holds and, moreover, these $\tilde K^\alpha$ also satisfy (\ref{ae}).

%We also have the following results. %following counterparts of Corollary~\ref{sol2} for
%the case of associativity equations (\ref{wdvv}).

Setting $K^\alpha=\eta^{\alpha\beta}\p F/\p x^\beta$, where $\eta^{\alpha\beta}$
is a symmetric nondegenerate constant matrix,
in Corollary~\ref{bt}
yields %the following result.
\begin{cor}\label{sol_oae}
Let the functions $F$
%=\check{F}(x^1,\dots,x^n)$
and $H^\alpha$
%=\check{H}^\alpha(x^1,\dots,x^n)$
satisfy the system %\vspace{-5mm}
\be\label{firint_wdvv_oae}
\eta^{\beta\nu}\eta^{\rho\kappa}\frac{\p^3 F}{\p x^\alpha\p x^\rho\p
x^\nu}\frac{\p^2 F}{\p x^\gamma\p x^\kappa}
=\ds\frac{\p^2 H^\beta}{\p x^\alpha\p x^\gamma}. %\alpha,\beta,\gamma=1,\dots,n,
\ee
Then $F$ automatically satisfies the
associativity equations (\ref{wdvv}) and $\tilde K^\alpha=H^\alpha$ automatically satisfy
the oriented associativity equations (\ref{ae}).\looseness=-1
\end{cor}

In complete analogy with the above,
(\ref{firint_wdvv_oae}) provides a B\"acklund
transformation relating the associativity equations (\ref{wdvv}) supplemented with the conditions
\be\label{firint_wdvv_oae_cor}
\eta^{\rho\kappa}\frac{\p^3 F}{\p x^\alpha\p x^\rho\p
x^\nu}\frac{\p^2 F}{\p x^\gamma\p x^\kappa}
=\eta^{\rho\kappa}\frac{\p^3 F}{\p x^\gamma\p x^\rho\p
x^\nu}\frac{\p^2 F}{\p x^\alpha\p x^\kappa}
\ee
which are necessary for (\ref{firint_wdvv_oae}) to hold,
and the oriented associativity equations (\ref{ae}) for $\tilde K^\alpha=H^\alpha$.

Note that the system (\ref{firint_wdvv_oae_cor}) was originally found and studied in \cite{m5, m6, m_geo3} (in
\cite{m5} it was referred to as a condition for compatible potential
deformation of a pair of the Frobenius algebras)
because this system plays an important role in classification of compatible
Hamiltonian structures of hydrodynamic type.
Moreover, in \cite{m7} (cf.\ also \cite{m_geo5}) it was proved that (\ref{firint_wdvv_oae_cor}) is also
equivalent to the condition of involutivity for a certain set of functionals
constructed from the function $F$ with respect to the constant homogeneous first-order
Poisson bracket of hydrodynamic type associated with the flat contravariant metric
$\eta^{\alpha\beta}$.\looseness=-1

%It would be interesting to find out whether one indeed could construct new
%classes of solutions for (\ref{ae}) and (\ref{wdvv}) using
%Corollaries~\ref{sol}--\ref{solwdvv}.

%Thus, the quantities
%\be\label{ints_wdvv}
%\eta^{ms}\frac{\p^3 F}{\p x^\alpha\p x^\gamma\p x^s}\frac{\p^2 F}{\p x^\nu\p x^\beta}
%\ee
%provide, in analogy with (\ref{ints}), a set of ``first integrals" for (\ref{wdvv}).
%It remains to be seen whether one could produce new classes of solutions for (\ref{wdvv})
%and, more broadly, for the WDVV equations, using the result of Corollary~\ref{solwdvv}.

%, and
% a general (smooth) solution of (\ref{firintwdvv})
%with {\em arbitrary} functions $G_\beta(x^1,\dots,x^n)$
%will simultaneously be a general solution for (\ref{wdvv}).

%It is easily seen that $\chi_k$ for $k\geq 2$ yield nonlocal symmetries for (\ref{wdvv}).

\section{Conclusions and open problems}
In the present paper we have found
infinite hierarchies of nonlocal higher symmetries
for the oriented and ``usual" associativity equations (\ref{ae})
and (\ref{wdvv}).
These symmetries can be employed %, in a standard way,
for producing new solutions from the known ones
and for constructing invariant solutions using the standard techniques as presented e.g.\ in
\cite{olv_eng2,vin2,kk}.\looseness=-1 %As a byproduct, we

%It would be interesting

%Moreover, it could be of interest to look
Moreover, it is natural to ask whether there exist
nonlocal symmetries and conservation laws for (\ref{ae}) (resp.\ for (\ref{wdvv}))
that depend on the nonlocal variables (\ref{wrec}),
(\ref{vdef_oae}) (resp.\ (\ref{vrec_oae}) with $K^\alpha=\eta^{\alpha\beta}\p F/\p x^\beta$)
in a more complicated fashion that
the symmetries found in Corollary~\ref{oae_new_sym2} (resp.\ Corollary \ref{oae_new_sym2wdvv}).
For instance, one could look for potential (in the sense of \cite{blu, rop} and references therein)
symmetries and conservation laws of (\ref{ae}) involving the nonlocal variables (\ref{wrec}) and (\ref{vdef_oae}).

The next steps to take include
elucidating the relationship among
the nonlocal symmetries of (\ref{wdvv})
from Corollary~\ref{sp1} and the symmetries found in \cite{mm} for
the generalized (in the sense of \cite{mor}) WDVV equations.
%, see e.g.\ \cite{mor} and
%references therein for the definition and properties of the latter.
The relationship (if any exists) of the flows (\ref{wdvvsymfl})
to the flows (5.15) from \cite{leur} could be of interest too.
Understanding the precise relationship of the symmetries
from Corollary~\ref{oae_new_sym2wdvv} to the tau-function and the B\"acklund transformations
for the WDVV equations from \cite{dub2} is yet another challenge.
%\looseness=-1
%
%Another interesting open problem is to find the counterparts of the
%nonlocal symmetries $\chi$, $\chi_k$ and $v_{k,\beta}$ for the case when $F$
%is subject to the conditions (\ref{etac}) and the quasihomogeneity conditions.
%
It would be also interesting to find recursion operators or master symmetries
for (\ref{ae}) and (\ref{wdvv}) that generate the hierarchies
from Corollaries~\ref{oae_new_sym2} and \ref{oae_new_sym2wdvv}.
\looseness=-1

Finally, it remains to be seen whether the Darboux-type transformation from
section 2 and the results from section 6 could indeed yield new solutions
for the oriented and ``usual" associativity equations.\looseness=-1

We intend
to address some of the above issues in our future work.\looseness=-1

\subsection*{Acknowledgments}

This research was supported in part by the Ministry of
Education, Youth and Sports of the Czech Republic (M\v{S}MT \v{C}R)
under grant MSM 4781305904, and by Silesian University in Opava under
grant IGS 9/2008.
\looseness=-1
%M.B. is pleased to acknowledge kind hospitality of
%the Mathematical Institute of Silesian University
%in Opava, where the present work was initiated.

%The author is very pleased
It is my great pleasure to thank E.V. Ferapontov, B.G. Konopelchenko,
M. Marvan, and M.V. Pavlov for reading the earlier drafts of the present manuscript
and making many useful suggestions,
and B. Dubrovin, M. Kontsevich, S. Shadrin, and I.A.B. Strachan
for stimulating discussions.
I am also pleased to thank the anonymous referee for suggesting a number of
useful improvements.

\looseness=-1

\end{document}